\documentclass[preprint,prb,showpacs,preprintnumbers,amsmath,amssymb]{revtex4}

\usepackage{graphicx}
\usepackage{color}
\usepackage{dcolumn}
\usepackage{bm}
\usepackage[latin1]{inputenc}

\begin{document}

\preprint{Accepted to PRB Vol. 79 Issue 5 (Feb-2009)/H. S. Ruiz and
A. Bad\'{\i}a-Maj\'{o}s}

\title{Nature of the Nodal Kink in Angle-Resolved Photoemission Spectra of Cuprate Superconductors}

\author{H. S. Ruiz$^{1,2}$}
\email[Electronic address: ]{hsruizr@unizar.es}
\author{A. Bad\'{\i}a-Maj\'{o}s$^{1}$}%
\affiliation{$^{1}$Departamento de F\'{\i}sica de la Materia
Condensada-I.C.M.A., Universidad de Zaragoza-C.S.I.C., Mar\'{\i}a de Luna 3, E-50018
Zaragoza, Spain\\$^{2}$Grupo de Superconductores No Homog\'{e}neos,
Universidad Nacional de Colombia, Bogot\'{a}, Colombia.}

\date{\today}

\begin{abstract}
The experimental finding of an ubiquitous kink in the nodal
direction of angle-resolved photoemission spectroscopies of
superconducting cuprates has been reproduced theoretically. Our
model is built upon the Migdal-Eliashberg theory for the electron
self-energy within the phonon-coupling scenario. Following this
perturbative approach, a numerical evaluation of the bare band
dispersion energy in terms of the electron-phonon coupling parameter
$\lambda$ allows a unified description of the nodal-kink effect. Our
study reveals that distinction between $\lambda$ and the technically
defined mass-enhancement parameter $\lambda^{*}$ is relevant for the
quantitative description of data, as well as for a meaningful
interpretation of previous studies.

A remarkable agreement between theory and experiment has been
achieved for different samples and at different doping levels. The
full energy spectrum is covered in the case of LSCO, Bi2212 and
overdoped Y123. In the case of underdoped Y123, the model applies to
the low energy region (close to Fermi level).

\end{abstract}


\maketitle

\section{Introduction}


High resolution angle-resolved photoemission spectroscopy (ARPES) is
nowadays considered as one of the most powerful methods for
obtaining detailed information about the electronic structure of
atoms, molecules, solids and surfaces.\cite{Hufner07} With the
advent of improved resolution, both in energy and momentum, ARPES
has provided key information on the electronic structure of high
temperature superconductors, including the band structure, Fermi
surface, superconducting gap, and
pseudogap.\cite{Zhou07,Damascelli03,Campuzano04} In this context,
its large impact on the development of many-body theories stems from
the fact that this technique provides a means of evaluating the
so-called electron self-energy, $\Sigma$. A wide and comprehensive
document on the more relevant aspects of the photoemission
spectroscopy is found in Ref.~\onlinecite{Hufner07}. As a
description of the spectroscopic techniques based on the detection
of photoemitted electrons is beyond the scope of this paper, below
we will only summarize some aspects that will be relevant in the
course of our discussion.

As regards the photoemission process, although it formally measures
a complicated nonlinear response function, it is helpful to notice
that the analysis of the optical excitation of the electron in the
bulk greatly simplifies within the
``sudden-approximation''.\cite{Randeria95,Hedin02} In brief, this
means that the photoemission process is supposed to occur {\em
suddenly}, with no post-collisional interaction between the
photoelectron and the system left behind.\cite{Damascelli03} In
particular, it is assumed that the excited state of the sample
(created by the ejection of the photo-electron) does not relax in
the time it takes for the photo-electron to reach the
detector.\cite{Zhou07} It can be shown that within the sudden
approximation using Fermi's Golden Rule for the transition rate, the
measured photocurrent density is basically proportional to the
spectral function of the occupied electronic states in the solid,
i.e.: $J_{\bf k}\propto A_{\bf k}(E)$. Eventually, and validated by
whether or not the spectra can be understood in terms of well
defined peaks representing poles in the spectral function, one may
connect $A_{\bf k}(E)$ to quasiparticle Green's function
$G(\textbf{k},E)=1/(E-\Sigma_{k}(E)-\varepsilon_{k})$, with
$\Sigma_{k}(E)$ defining the self-energy and $\varepsilon_{k}$ the
bare band dispersion. In fact, $A(\textbf{k},\omega)=-(1/\pi)$Im$
G(\textbf{k},\omega+i0^{+})$. Beyond the sudden approximation, one
would have to take into account the screening of the photoelectron
by the rest of system, and the photoemission process could be
described by the generalized golden rule formula, i.e, a
three-particle correlation function.\cite{Hedin02} For our purposes,
it is important mention the evidence that the sudden approximation
is justified for the cuprate superconductors even at low photon
energies.\cite{Randeria95,Koralek06} In the end, the suitability of
the approximations invoked, will be justified by the agreement
between the theory and the experimental observations for the wide
set of data.


Interactions involving a low-energy excitation appear as a sudden
change in the electron energy dispersion near the Fermi level
($E_{F}$), known as \emph{kink}.\cite{Zhou07} This feature has been
observed in various CSC and at different doping levels both along
the nodal
\cite{Lanzara01,Zhou03,Zhou05,Kordyuk06,Zhou02,Xiao07,Takahashi07,Gweon04,Douglas07,Johnson01,Borisenko06}
and antinodal \cite{Zhou02,Xiao07,Takahashi07,Gweon04,Douglas07}
directions, revealing diverse and controversial behaviors commonly
interpreted in terms of the coupling of electrons to a certain kind
of excitation. Recall that the antinodal direction denotes the
($\pi$, 0) region in the Brillouin zone where the \emph{d}-wave
superconducting gap has a maximum. This fact extraordinarily
complicates the theoretical interpretation of the kink, due to the
anisotropic character beyond the s-channel.\cite{Zhou07} More
advantageous is the nodal direction that corresponds to the
(0,0)-($\pi$,$\pi$) direction in the Brillouin zone, where the
\emph{d}-wave superconducting gap is zero. Multiple measurements
have been realized along this direction in several CSC both in the
normal and superconducting (SC) state. Such measurements show a kink
in a similar energy scale (in the range of 48-78 meV) and are
present over an entire doping range, and for temperatures well below
and above T$_{c}$.


To the moment, there is no consensus on the origin and behavior of
the kink in the CSC and its influence on the bosonic coupling
mechanism that leads to SC state. Even more, there is no model
allowing to reproduce the kink effect in different materials and/or
different doping levels. For this reason, a theoretical description
of the kink effect in various CSC and at different doping levels is
quite desirable. In this article, with the aim of obtaining a
correct dressed electron band dispersion relation which can
reproduce a wide number of experimental spectroscopies, we suggest a
model based on the Migdal-Eliashberg (ME) approach for the numerical
determination of the electron self-energy $\Sigma$. As a main
result, we emphasize the importance of considering the proper
distinction between the mass-enhancement parameter $\lambda^{*}$ and
the electron-phonon coupling parameter $\lambda$.


The essence of the Eliashberg theory is a perturbative scheme that
allows to deal with strong electron-phonon coupling effects. It
relies on the Migdal approach for the theory of metals, that is
posed in terms of thermal Green's functions $G$ for systems of many
interacting particles described by a hamiltonian $H$ within the
Fermi liquid picture.\cite{Allen82,Ruiz08} Eventually, one is lead
to the definition $G^{-1}=G_{0}^{-1}-\Sigma$, where $G_{0}$ stands
for the non-interacting electrons, whereas $\Sigma$ is the above
referred self-energy function which is a measure of the perturbation
introduced to the bare Hamiltonian by the interactions. Remarkably,
the averaging procedures for obtaining $\Sigma$ happen to be
expressed in terms of experimentally accessible data ($\alpha^{2}F$)
convoluted with well known mathematical functions ($R$), i.e.:
$\Sigma \propto \alpha^{2}F \star R$. $\alpha^{2}F$ will be derived
from inelastic neutron scattering experiments and $R$ defined in
terms of the so-called digamma functions. In this work, the
comparative study of a wide set of experimental results has allowed
an empirical extension of the commonly considered physical scenario
when analyzing ARPES data.


The paper is organized as follows. First, in Sec.\ref{Sec:sigma}, we
will give details about the relevant quantities and theoretical
treatment (ME approach) to be used. The physical interpretation of
the underlying approximations is also focused. In
Sec.\ref{Sec:analysis}, the analysis of the ARPES data according to
our proposal is done. Comparison with previous published material
will be emphasized. Finally, Sec.\ref{Sec:discussion} is devoted to
discuss our results. The relevance of the electron-phonon coupling
mechanism for the interpretation of the nodal-kink of ARPES
experiments in CSC will be concluded.

\section{The electron self-energy $\Sigma$}
\label{Sec:sigma}


In ARPES the dressed electronic dispersion relation is denoted by
$E_{\textbf{k}}$.\cite{Zhou02} This quantity characterizes the
charge carriers as quasiparticles that are formed when the electrons
are {\em dressed} with excitations. $E_{\textbf{k}}$ has been
commonly related to the bare band dispersion
$\varepsilon_{\textbf{k}}$ through the real part of the electron
self-energy by
$E_{\textbf{k}}=\varepsilon_{\textbf{k}}+Re\Sigma(E_{k})$. Within
the Eliashberg theory, the electron-phonon interaction (EPI)
self-energy may be obtained from the real part of the expression
\cite{Allen82}
%
%
\begin{widetext}
\begin{equation}
\Sigma(\omega+i0^{+})=\int_{0}^{\infty}d\nu\alpha^{2}F(\nu)\left\{-2\pi
i\left[N(\nu)+\frac{1}{2}\right]+\psi\left(\frac{1}{2}+i\frac{\nu-\omega}{2\pi
T}\right)-\psi\left(\frac{1}{2}-i\frac{\nu+\omega}{2\pi
T}\right)\right\}\, ,\label{eq:one}
\end{equation}
\end{widetext}
%
%
valid for the whole range of temperatures $T$, frequencies $\nu$,
and energies $\omega$. Here, $\psi(z)$ are the so-called digamma
functions with complex argument and $\alpha^{2}F(\nu)$ defines the
important EPI spectral density which measures the effectiveness of
the phonons of frequency $\nu$ in the scattering of electrons from
any state to any other state on the Fermi surface. More
specifically, being interested in the nodal direction ARPES
experiments, which are not influenced by the anisotropy of the
superconducting gap, we will refer to a (non-directional) isotropic
quasiparticle spectral density, defined as the double average over
the Fermi surface of the spectral density
$\alpha^{2}F(\textbf{k},\textbf{k'},\nu)$, i.e.:
%
%
\begin{eqnarray}
    \alpha^{2}F(\nu)=\frac{1}{N(0)}\sum_{\textbf{kk'},j} \mid
    g_{\textbf{kk'}}^{j}\mid^{2}\delta(\nu-\nu_{\textbf{k}-\textbf{k'}}^{j})
    \delta(\varepsilon_{k})\delta(\varepsilon_{k'}) \label{eq:two}
\end{eqnarray}
%
%
where,
$g_{\textbf{k}\textbf{k}'}^{j}=[\hbar/2M\nu^{j}_{\textbf{k}'\textbf{k}}]^{1/2}\langle
\textbf{k}|\hat{\epsilon}^{\; j}_{\textbf{k}'\textbf{k}}\cdot\nabla
V|\textbf{k}'\rangle$ defines the EPI-matrix element for electron
scattering from $\textbf{k}$ to $\textbf{k'}$ with a phonon of
frequency $\nu_{\textbf{k}-\textbf{k'}}^{j}$ (j is a branch index).
$M$ stands for the ion mass, $V$ is the crystal potential,
$\hat{\epsilon}^{\; j}_{\textbf{k}'\textbf{k}}$ is the polarization
vector, and $N(0)=\sum_{k}\delta(\varepsilon_{k})$ represents the
single-spin electronic density of states at the Fermi surface. As
usual $\delta(x)$ denotes the Dirac's delta function evaluated at
$x'=0$. Owing to the intrinsic complexity for evaluating
$\alpha^{2}F(\nu)$ for a given material from first principles, in
this work we have adopted a more empirical point of view, which
relies on auxiliary experimental data.

\begin{figure}
\includegraphics[width=0.6\textwidth]{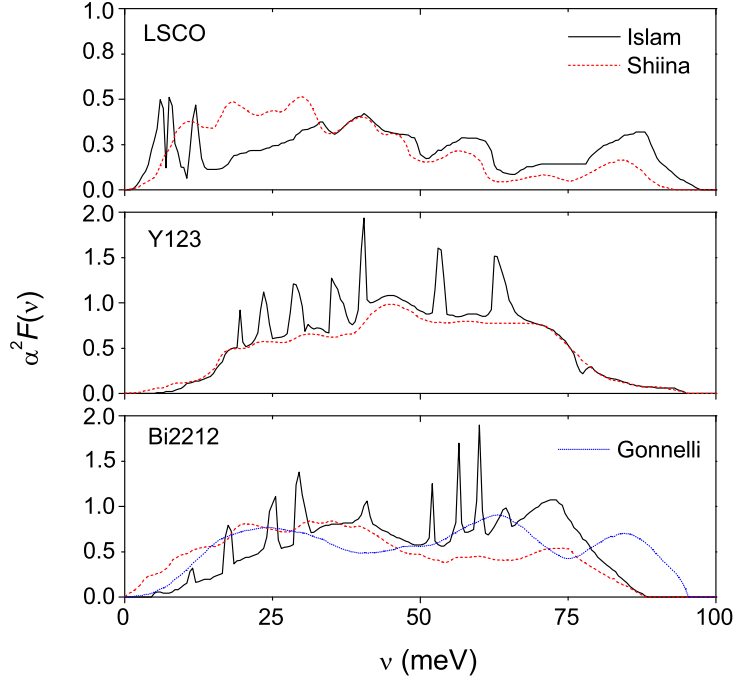}
\caption{\label{Fig:1} (color online) EPI spectral density
$\alpha^2\emph{F}(\nu)$ for LSCO, Y123 and Bi2212. The black solid
lines correspond to the method of Ref.~\onlinecite{Islam00}. The
other lines are shown for comparison employing the methods of
Ref.~\onlinecite{Shiina90} in all materials (dashed lines), and from
Ref.~\onlinecite{Gonnelli98} in Bi2212 (dotted lines).}
\end{figure}

Taking into consideration the inherent existence of phonons in the
CSC, and for simplicity, we have chosen the model by Islam \& Islam
\cite{Islam00} for the extraction of the EPI spectral function
$\alpha^{2}F(\nu)$ through the phonon density states obtained from
inelastic neutron scattering in LSCO,\cite{Renker87}
Bi2212,\cite{Renker89} and Y123.\cite{Renker88} The corresponding
spectral densities are shown in Fig.~\ref{Fig:1}, in comparison with
the spectral densities obtained by Shiina and
Nakamura,\cite{Shiina90} and Gonnelli $et$ $al.$\cite{Gonnelli98} We
emphasize that very similar results are found under the use of any
of these densities.
%

%

At this point, we should comment that other choices of the spectral
density are possible, but have been left aside. For instance, the
absence of the magnetic-resonance mode in LSCO and its appearance
only below $T_{c}$ in some CSC are not consistent with the idea that
the nodal kink has a magnetic origin. \cite{Zhou07} Along the same
line, the reported absence of the magnetic-resonance mode in Bi2212
at a doping level of 0.23 without the disappearance of the
superconducting state\cite{Hwang04} ($T_{c}\approx 55 K$) reveals
that the magnetic-resonance mode cannot be directly related to the
spectral density involved in Eq.~(\ref{eq:one}). Therefore, we have
preferred the spectral densities oriented to phonons.


An important ingredient of our model is the commonly used
dimensionless electron-phonon coupling parameter, defined in terms
of the EPI spectral function by $\lambda\equiv
2\int_{0}^{\infty}d\nu\,\alpha^{2}F(\nu )/\nu$. This quantity is not
to be straightforwardly identified with the mass-enhancement
parameter $\lambda^{*}$. As it will be shown below, coincidence is
only warranted under certain limits. A further relevant feature to
mention is that $|g_{\textbf{k}\textbf{k}'}^{j}|^{2}$ and as a
consequence $\lambda$ are inversely proportional to the number of
charge carriers contributed by each atom of the crystal to the
bosonic coupling mode. Therefore, an increase in the doping level,
which causes an increment in the hole concentration of the $CuO_{2}$
plane must be reflected in a reduction of $\lambda$ as we will see
in the analysis of the kink structure. Furthermore, recalling the
outstanding feature of the theory of metals, that
$|g_{\textbf{k}\textbf{k}'}^{j}|^{2}$ vanishes linearly with
$|\textbf{k}-\textbf{k}'|$ when $|\textbf{k}-\textbf{k}'| \ll
{k}_{F}$,\cite{Ashcroft76} one would expect a {\em linear}
disappearance of the coupling effect that gives rise to the nodal
kink in the vicinity of the Fermi surface. On the other hand,
inspired by recent results on the universality of the nodal Fermi
velocity $v_{F<}$ (at low energies) in certain cuprates, a prominent
role of this quantity is also expected.

\subsection{Green's function formalism}


Some words are due, considering Eq.(\ref{eq:one}) and the relation
between the dressed and bare energies. Within the arguments
customarily used for analyzing ARPES data, one finds the low energy
approximation (close to the Fermi surface)
$E_{k}\approx\varepsilon_{k}/(1+\lambda)$, with $\lambda$ the
electron-phonon coupling parameter defined above. However, one
should recall that such expression is an asymptotic form of the more
correct $E_{k}\approx\varepsilon_{k}/(1+\lambda^{*})$ with
$\lambda^{*}$ the true mass-enhancement parameter and $\lambda$ an
appropriate limiting value. As a central result of our work,
distinction between them is essential for the overall description of
available data. Let us show how this arises.


Following the thermal Green's function formalism, we assume that the
electron-phonon interaction is introduced by the relation
%
%
\begin{equation}
G^{-1}(k,i\omega_{n})=G_{0}^{-1}(k,i\omega_{n})-\Sigma(k,i\omega_{n})
\label{eq:three}
\end{equation}
with $G_{0}^{-1}$ related to the bare electron energy and
$i\omega_{n}$ standing for the ``imaginary Matsubara
frequencies''.\cite{Allen78}
%


We recall that, technically, the bare electron band energy
$\varepsilon_{k}$ is determined by the poles of the Green function
$G(k,i\omega_{n})$, or the zeros of $G^{-1}(k,i\omega_{n})$ at the
$i\omega_{n}$ frequencies.\cite{Allen78} On the other hand, it is
known that additional dynamical information is contained in the
analytic continuation $G(k,\omega+i0^{+})$ to points just above the
real frequency axis, known as the ``retarded'' Green's function. One
is therefore led to continue the electronic self-energy
$\Sigma(k,\omega+i0^{+})$ analytically by
$\Sigma(k\omega+i0^{+})\equiv
\Sigma_{1}(k,\omega)+i\Sigma_{2}(k,\omega)$. Now, suppose that a
pole occurs near $\omega=0$ so, one gets
%
%
\begin{eqnarray}
&G^{-1}(k,\omega+i0^{+})=\omega-\varepsilon_{k}-\Sigma_{1}(k,\omega)-i\Sigma_{2}(k,\omega)\\
\label{eq:four}&\simeq
\omega\left(1-\left.\frac{\partial\Sigma_{1}(k,\omega)}{\partial\omega}\right|_{\omega=0}\right)-
\left[\varepsilon_{k}+\Sigma_{1}(k,0)\right]-i\Sigma_{2}(k,\omega)\nonumber
\; .%
\end{eqnarray}%
Then, the pole of $G$ occurs at a frequency $\omega_{0}$ given by
$\omega_{0}=E_{k}-i/2\tau_{k}$, with
$\tau_{k}^{-1}=-2\left(1-\partial_{\omega}\Sigma_{1}\right)^{-1}\Sigma_{2}(k,E_{k})$
and
%
%
\begin{equation}
E_{k}=(1-\partial_{\omega}\Sigma_{1})^{-1}\left[\varepsilon_{k}+\Sigma_{1}(k,0)\right]
\; .\label{eq:five}%
\end{equation}
Here
$\lambda_{k}^{*}\equiv\left.-\partial_{\omega}\Sigma_{1}\right|_{\omega=0}$
is the technically defined mass-enhancement parameter.\cite{Allen82}


We want to emphasize that replacement of $\lambda^{*}$ by $\lambda$
in Eq.(\ref{eq:five}) is only warranted for states $k$ at or very
close to the Fermi surface when the ME approach for the self-energy
[Eq.(\ref{eq:one})] has been employed at the low temperature limit
($T\to 0$). Again, owing to the difficulties for evaluating
$\Sigma_{1}$ beyond the Migdal approximation, our position in this
paper has been to obtain $\lambda^{*}$ through the systematic
evaluation of ARPES data, as shown in
Sec.\ref{subsec:phenomenological}. As a further detail, related to
the limitations introduced by the use of the electron-phonon
coupling parameter, one should mention that, long before the advent
of the high $T_{c}$ superconductivity, Ashcroft and Wilkins showed
that in some simple metals the single parameter $\lambda$ is
insufficient to determine a number of thermodynamic properties such
as the specific heat. \cite{Ashcroft65}


In the following, we will introduce the simplest correction
possible, for dealing with the correct mass-enhancement parameter.
Noteworthily, it will be shown that a phenomenological linear
relation, i.e.: $\lambda^{*}\simeq\delta\lambda$ suffices for the
interpolation of experiments and numerical data. The physical
interpretation of the parameter $\delta$ will be discussed within
the following section (\ref{subsec:phenomenological}). Just from the
technical side, we want to comment that the mathematical material
within this section has been developed using the convention of
positive energy bands relative to the Fermi surface. It is apparent
that the contrary selection can also be done, changing signs within
intermediate expressions but the same final results, i.e.:
$E_{k}=(1+\partial_{\omega}\Sigma_{1})^{-1}\left[\varepsilon_{k}-\Sigma_{1}(k,0)\right]$,
with $\varepsilon_{k}$ defining negative energy bands.

\subsection{Phenomenological dispersion relation}\label{subsec:phenomenological}


Let us make some final remarks before moving onto the application of
the above ideas to the analysis of the ARPES data. Let us start by
recalling that the bare electron band energy $\varepsilon_k$ is not
directly available from the experiments. Instead, the electron
momentum dispersion curve $E_{k}(k-k_{F})$ may be measured. However,
it has been noted that the relation between the dressed and bare
energies is a central property as related to the kink structure. In
fact, based on the commonly used equation
$E_{k}=\varepsilon_{k}+Re\{\Sigma(E_{k})\}$ one can consider that
$\varepsilon_k$ implicitly depends on $E_k$ through the boson
coupling parameter $\lambda$.\cite{Schachinger03,Giustino08} This
fact, along with some ansatz for the ARPES ``bare'' dispersion
allows to obtain $\lambda$ as a unique constrained parameter that
better fits the observed kink structure. However, the indiscriminate
proposal of dispersion relations could considerably under or
overestimate the average renormalization because implicit
approximations are used as indicated in the previous section.
Furthermore, we stress that, the use of the same ansatz on different
cuprates or even for a definite material with slight variations in
the doping level is not warranted. In this work, with the aim of
finding a widespread renormalization function that describes the
behavior of the kink in different CSC and for different doping
levels, we have carried out an exhaustive study on the incidence of
the EPI coupling parameter in the appearance of kink-dispersion.
Based on an interpolation scheme between the numerical behavior of
the relation $E_{k}=\varepsilon_{k}+Re\{\Sigma(E_{k})\}$ and the
experimental data, as a central result, we have encountered that the
practical totality of data are accurately reproduced by a universal
dispersion relation of the kind
%
%
\begin{equation}
{k}-{k}_{F}=\frac{\varepsilon_{k}}{v_{F_{<}}}\left(1-\frac{\varpi}{\omega_{\rm
log}}\lambda\right)\, ,\label{eq:six}
\end{equation}
%
with $v_{F_{<}}$ the Fermi velocity at low-energies, $\omega_{\rm
log}$ the so-called logarithmic frequency as introduced by
Allen,\cite{Allen75} and $\varpi$ the (only) free parameter required
for incorporating the specific renormalization for each
superconductor. Note that the constant frequency $\omega_{\rm log}$
(properly defining the corresponding spectral densities
$\alpha^{2}F(\nu)$ for each $\lambda$) has been introduced only for
scaling purposes, just with the aim of reducing the scattering of
numerical values in dealing with different samples. In this sense,
$\varpi$ is a mere mathematical instrument. Thus, our numerical
program is as follows: (i) $v_{F_{<}}$ is determined from the
momentum dispersion curve in the ARPES measurements [$\sim $ 1.4
eV{\AA} - 2.2 eV{\AA}], (ii) $\omega_{\rm log}$ is evaluated from
the spectral densities shown in Fig.\ref{Fig:1} through the
definition $\omega_{\rm log}\equiv
exp\{(2/\lambda)\int_{0}^{\infty}\ln(\nu)[{\alpha^{2}F(\nu)}/{\nu}]d\nu\}$
(we get $\omega_{\rm log}^{LSCO}\simeq 16.1455\, meV$, $\omega_{\rm
log}^{Y123}\simeq 35.5900\, meV$ and $\omega_{\rm
log}^{Bi2212}\simeq 33.8984\, meV$ respectively), (iii)
$\varepsilon_{k}$ is numerically determined from the relation
$E_{k}=\varepsilon_{k}+Re\{\Sigma(E_{k})\}$, and (iv) correlation is
established between theory and experiment by the application of
Eq.(\ref{eq:six}).


From the physical point of view, our empirical {\em ansatz}
[Eq.(\ref{eq:six})] may interpreted as follows. Let us assume that
the involved quantities are not far from their values at the Fermi
level, and start with $\varepsilon_k$ replaced by
$E_{k}-\Sigma_{1}(E_k)$, i.e.
%
%
\begin{equation}
E_{k}-\Sigma_{1}(E_k)=\left({k}-{k}_{F}\right)v_{F_{<}}\left(1-\delta\lambda\right)^{-1}\,
,\label{eq:seven}
\end{equation}
where the definition $\delta\equiv{\varpi}/{\omega_{\rm log}}$ has
been used. Now, let us take derivatives respect to $E_k$ and
evaluate for $E_k\to 0$. One gets
%
%
\begin{equation}
1+\lambda^{*}=\left.\frac{\partial (k-k_F)}{\partial
E_k}\right|_{E_{k}=0} v_{F<}(1-\delta\lambda)^{-1}\,
,\label{eq:eight}
\end{equation}
and recalling that $v_{F<}$ is obtained as the slope of the lower
part of the momentum dispersion curve,\cite{Zhou03} this equation
leads to $1+\lambda^{*}=(1-\delta\lambda)^{-1}$. Thus, a physical
interpretation of the fit parameter $\delta$ is obtained, i.e.:
$\delta = (\lambda^{*}/\lambda)/(1+\lambda^{*})$. To the lowest
order, the dimensionless parameter $\delta$ is basically the ratio
between the defined mass-enhancement and phonon-coupling parameters
$\delta\approx\lambda^{*}/\lambda$. Outstandingly, it will be shown
that this fact reassembles the differences obtained by tight-binding
Hamiltonian models \cite{Weber87,Weber88} and the
``density-functional'' band theories \cite{Giustino08,Heid08}.
Recall that, in principle, the density-functional theory
\cite{Kohn66} gives a correct ground-state energy, but the bands do
not necessarily fit the quasi-particle band structure used to
describe low-lying excitations. As it will be seen below,
predictions from both types of models may be reconciled appealing to
the differences between $\lambda$ and $\lambda^{*}$.


\begin{figure}
\includegraphics[height=0.6\textwidth]{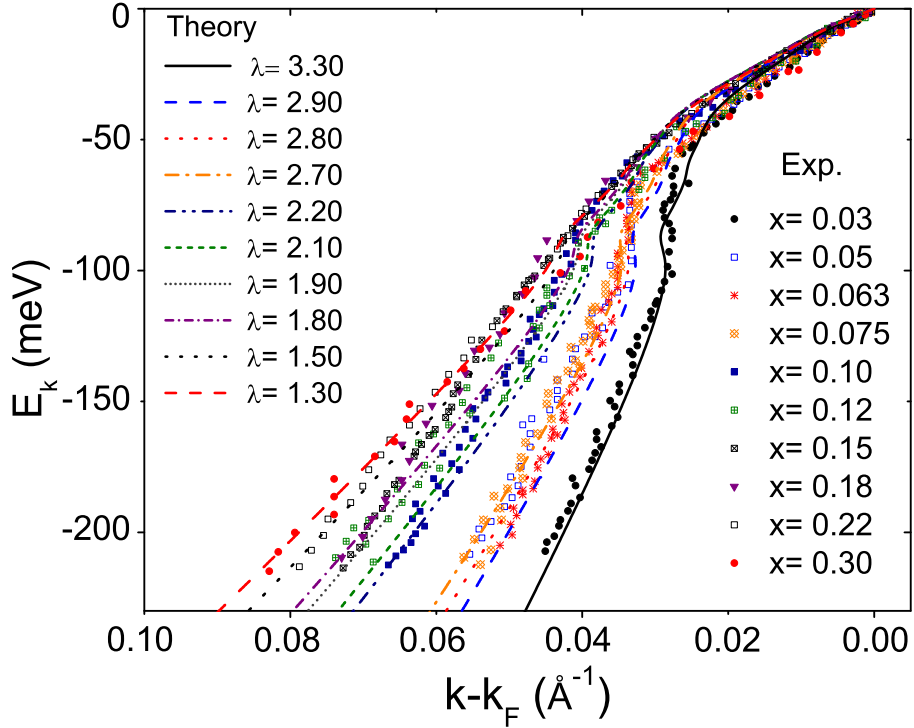}
\caption{\label{Fig:2} (color online) The renormalized energy
$E_{k}$ as a function of momentum $k-k_{F}$ for
$La_{2-x}Sr_{x}CuO_{4}$ with various doping levels between 0.03
(right curve) and about 0.30 (left curve), all measured at a
temperature of $20K$ along the nodal direction. Experimental data
are taken from Ref.~\onlinecite{Zhou03}. The theoretical curves are
labeled according to the best fit values for $\lambda$.}
\end{figure}
%
%

\begin{table}
\caption{\label{tab:table1} Values of the boson-coupling parameter
$\lambda$ and the corresponding mass-enhancement parameter
$\lambda^{*}$ obtained from the analysis of ARPES data at several
doping levels of $La_{2-x}Sr_{x}CuO_{4}$. $\lambda^{*}$ has been
obtained by means of Eq.~(\ref{eq:six}) to the lowest order
$\lambda^{*}\approx\delta\lambda$ (in this case $\delta=0.185$). The
predicted superconducting transition temperatures $T_{c}$ are also
shown. Our results are presented in contrast with other models
available in the literature.}
\begin{ruledtabular}
\begin{tabular}{cccccc}
SC&$x$&Ref.&$\lambda$&$\lambda^{*}$&$T_{c}(\lambda)$  [K]\\
\hline
LSCO & 0.03  & This\footnotemark[1] & 3.30 & 0.61 & -- \\
     & 0.05  &      & 2.90 & 0.54 & -- \\
     & 0.063 &      & 2.80 & 0.52 & -- \\
     & 0.075 &      & 2.70 & 0.50 & -- \\
     & 0.10  &      & 2.20 & 0.41 & 42.10 \\
     & 0.12  &      & 2.10 & 0.39 & 40.77 \\
     & 0.15  &      & 1.90 & 0.35 & 37.83 \\
     & 0.18  &      & 1.80 & 0.33 & 36.19 \\
     & 0.22  &      & 1.50 & 0.28 & 30.47 \\
     & 0.30  &      & 1.30 & 0.24 & -- \\
\hline
     & 0.1-0.2 & [\onlinecite{Weber87}]\footnotemark[2] & 2-2.5 & -- & 30 - 40 \\
%
\hline
     & -- & [\onlinecite{Shiina90}]\footnotemark[2] & 1.78 & -- & 40.6 \\
%
\hline
     & 0.15  & [\onlinecite{Giustino08}]\footnotemark[3] & 1 - 1.32 & 0.14 - 0.22 & -- \\
     & 0.22  & [\onlinecite{Giustino08}]\footnotemark[3] & 0.75 - 0.99 & 0.14 - 0.20 & -- \\
%
\end{tabular}
\end{ruledtabular}
\footnotetext[1]{We allow a margin of error in $\lambda$ of $\sim
\pm 0.3$ as related to the numerical interpolation procedure between theory and experiment.}
\footnotetext[2]{The $\lambda$ values reported
in that reference were obtained so as to fit $T_{c}$ at the indicated values.}
\footnotetext[3]{In Ref.~\onlinecite{Giustino08} the electronic structure of LSCO has been
calculated employing a generalized gradient approximation to density
functional theory and used to determine $\lambda$.}
\end{table}

\section{Analysis of ARPES data}
\label{Sec:analysis}

\begin{figure}
\includegraphics[height=0.6\textwidth]{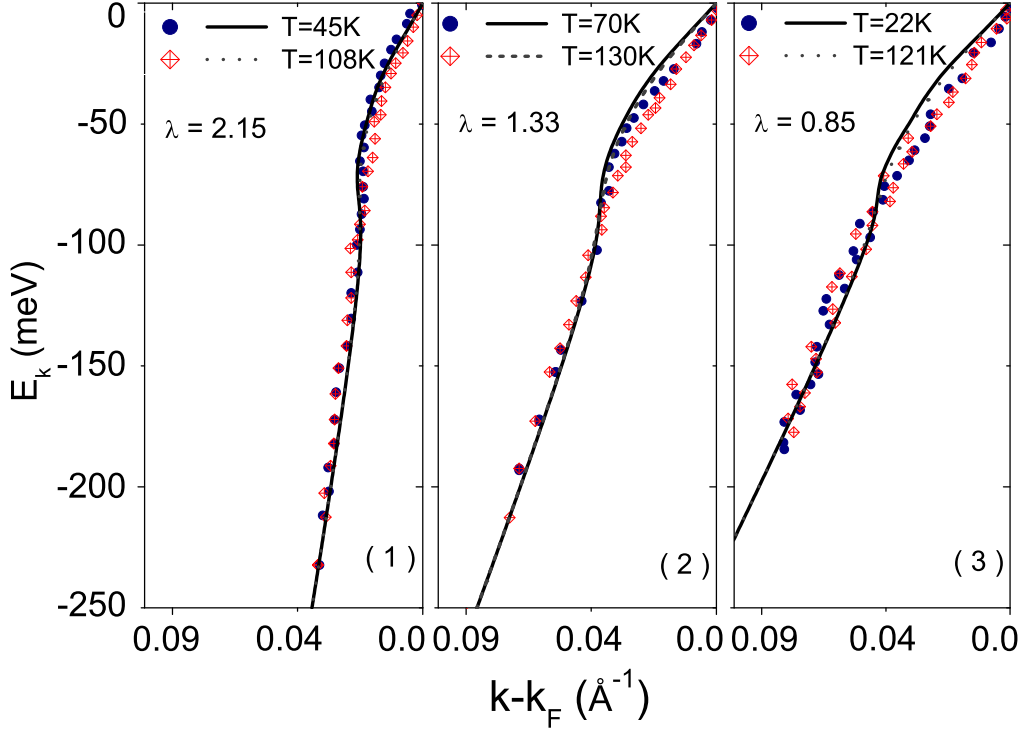}
\caption{\label{Fig:3} (color online)  Same as in Fig. [2] but in
samples of $Bi_{2}Sr_{2}CaCu_{2}O_{8+x}$ with (left to
right):$x=0.12$ (underdoped), $x=0.16$ (optimally doped), and
$x=0.21$ (overdoped). The red diamonds and blue circles correspond
to the experimental data of Ref.~\onlinecite{Johnson01}. The
continuous lines correspond to our theoretical curves.}
\end{figure}
\begin{figure}
\includegraphics[height=0.6\textwidth]{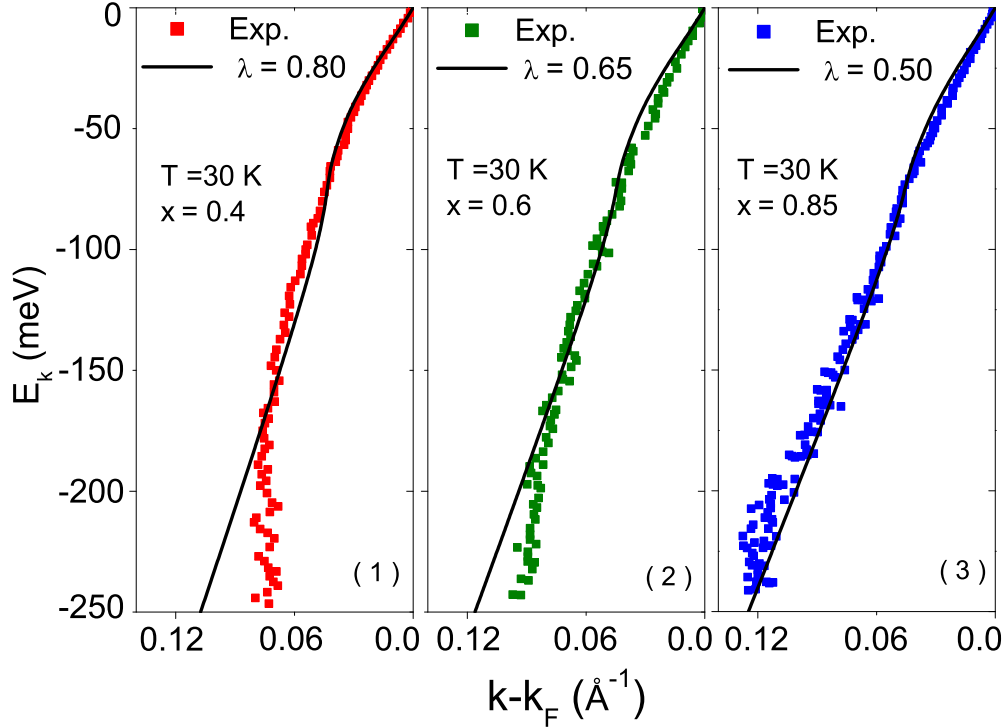}
\caption{\label{Fig:4} (color online) Same as in Fig. [2] but in
samples of $YBa_{2}Cu_{3}O_{6+x}$ with (left to right): $x=0.4$
(underdoped), $x=0.6$ (underdoped), and $x=0.85$ (overdoped). The
solid squares correspond to the experimental data of
Ref.~\onlinecite{Borisenko06}. The lines correspond to our
theoretical curves. All curves have been obtained at 30K.}
\end{figure}

Below, we present the application of our theoretical analysis for a
wide set of experimental curves available in the literature. The
main facts are shown in figures~\ref{Fig:2}-~\ref{Fig:4}, and
summarized in {tables~\ref{tab:table1}-\ref{tab:table3}.}
%
%
\paragraph{Results for LSCO.--} In Fig.~\ref{Fig:2} we show the results found in
$La_{2-x}Sr_{x}CuO_{4}$ covering the doping range ($0<x\leq0.3$).
Remarkably, within this range, the physical properties span over the
insulating, superconducting, and overdoped non-superconducting metal
behavior. Superconducting transition temperatures $T_{c}$ in the
interval of 30-40K have been observed by Bednorz and
M\"{u}ller~\cite{Bednorz86} and others~\cite{Uchida87,Dietrich87}.
For the application of Eq.(\ref{eq:six}), here, we have considered
$v_{F_{<}}=2 eV\cdot${\AA} as related to the experimental results of
Refs. \onlinecite{Zhou07,Lanzara01,Zhou03,Zhou05}. On the other
hand, the best fit of the whole set of experimental data has been
obtained for $\delta=0.185$, and the derived $\lambda^{*}$ values
are shown in the Table~\ref{tab:table1}. For comparison, recall that
values of ``$\lambda=2-2.5$'' in the range $0.2>x>0.1$ were reported
in the Ref.~\onlinecite{Weber87} by Weber. In that case they were
obtained within the framework of the nonorthogonal-tight-binding
theory of lattice dynamics, based on the energy band results of
Mattheiss~\cite{Mattheiss87} and corresponding to the range of the
EPI coupling parameter $\lambda$ predicted in our model. It must be
emphasized that in the case of Ref.~\onlinecite{Weber87}, $\lambda$
was obtained in agreement with the observed $T_c$ values in LSCO.
Moderate discrepancies between the predictions of our
phenomenological model and the analysis of
Refs.~\onlinecite{Weber87} and \onlinecite{Weber88} may be ascribed
to some uncertainty in the experimental spectral densities. As
regards the critical temperatures, we have calculated them based on
the celebrated McMillan's equation,\cite{McMillan68}
$T_{c}=(\omega_{log}/1.2)exp[-1.04(1+\lambda)/(\lambda-\mu^{*}(1+0.62\lambda))]$
(see table~\ref{tab:table1}).  In all the calculations, the
Coulomb's pseudopotencial was given a typical value, $\mu^{*}=0.13$.
It is essential to be aware that, there is no small parameter which
enables a satisfactory perturbation theory to be constructed for the
Coulomb interaction between electrons. Thus Coulomb contributions to
the electron self-energy $\Sigma$ cannot be reliably
calculated.\cite{Allen75} Fortunately this is not a serious problem
in superconductivity because a reasonable assumption is to consider
that the large normal-state Coulomb effects contained in the Coulomb
self-energy are already included in the bare band structure
$\varepsilon_{k}$. The remaining off-diagonal terms of the
superconducting components of the Coulomb self-energy turn out to
have only a small effect on superconductivity, which is treated
phenomenologically.\cite{Allen82} One can see that the consideration
of the electron-phonon interaction in LSCO strongly suggests that
the high $T_{c}$ values can be caused by conventional
electron-phonon coupling, in agreement with the conclusion of
Weber.~\cite{Weber87}


From a different perspective, in a recent publication, Giustino and
co-workers\cite{Giustino08} have calculated the electronic structure
of LSCO employing a generalized gradient approximation to density
functional theory (DFT). These authors have extracted the
``$\lambda$'' parameter by measuring the gradients of both the
theoretical and the experimental~\cite{Lanzara01} self-energy data
within the low energy limit ($0-50meV$). Their procedure yields
$\lambda_{expt}=1.00-1.32$ for the optimally doped sample ($x=0.15$)
at 20K and $\lambda_{expt}=0.75-0.99$ for the overdoped sample
($x=0.22$) while the theoretical results $\lambda_{th}=0.14-0.22$ at
optimal doping and $\lambda_{th}=0.14-0.20$ in the overdoped regime.
It is concluded that theoretical values noticeably underestimate the
experiments and, thus, that the electron-phonon interaction is
unlikely to be relevant in LSCO. From our view, considering that the
theoretical value is calculated from the gradient of the DFT
self-energy, $\lambda_{th}$ is basically to be identified with
$\lambda^{*}$, while $\lambda_{expt}$ matches the electron-phonon
coupling $\lambda$ involved in the standard Migdal formalism
analysis of experiments.
%

%
\begin{table}
\caption{\label{tab:table2} Same as table~\ref{tab:table1} but for
the case of $Bi_{2}Sr_{2}CaCu_{2}O_{8+x}$.Here, several values of
$\lambda$ are not related to ARPES data, but just introduced for
comparison. In this case, $\delta=0.354$}
\begin{ruledtabular}
\begin{tabular}{cccccc}
SC&$x$&Ref.&$\lambda$&$\lambda^{*}$&$T_{c}(\lambda)$  [K]\\
\hline
Bi2212 & 0.12 & This\footnotemark[1] & 2.15 & 0.76 & 64.81 \\
       & 0.16 &      & 1.33 & 0.47 & 42.45 \\
       & 0.21 &      & 0.85 & 0.30 & 19.93 \\
\hline
       &  --  & [\onlinecite{Shiina90}]\footnotemark[2] & 3.28 & -- & 85 \\
       &  --  & This  & 3.28 & 1.16 & 81.66 \\
\hline
       &  --  & [\onlinecite{Gonnelli98}]\footnotemark[2] & 3.34 & 1.05 & 93 \\
       &  --  & This  & 3.34 & 1.18 & 82.40 \\
\hline
       & 0.16 & [\onlinecite{Kordyuk06}]\footnotemark[3] & $\sim 1.28$ & $\sim 0.43$ & -- \\
%
\end{tabular}
\end{ruledtabular}
\footnotetext[1]{We allow a margin of error in $\lambda$ of $\sim
\pm 0.3$ as related to the numerical interpolation procedure between theory and experiment.}
\footnotetext[2]{The $\lambda$ values reported
in that reference were obtained so as to fit $T_{c}$ at the indicated values.}
\footnotetext[3]{In Ref.~\onlinecite{Kordyuk06} two channels are
defined for $\lambda$. $\lambda_{1}=0.43\pm0.02$ corresponds to the
``\emph{primary}'' channel (close to Fermi level) and is free from
normalization effects. $\lambda_{2}$ is obtained from the
Kramers-Kronig transformation and the experiment.\cite{Kordyuk05} In
this sense, and within the notation employed in the current work, we
obtain, $\lambda^{*}=\lambda_{1}$, and $\lambda\simeq0.85+0.43$.}
\end{table}


\paragraph{Results for BSCCO.--} In Fig.~\ref{Fig:3} we display the results found in samples of
Bi2212. Each case has been studied with temperatures both in the
normal and superconducting state. The experimental data were taken
from the work by Johnson \emph{et al.}\cite{Johnson01} In this case,
we have used $v_{F_{<}}=1.6eV\cdot${\AA} as a value consistent with
the experimental results of
Refs.~\onlinecite{Lanzara01,Kordyuk06,Zhou02,Johnson01}. The best
fit with experimental data has been found for $\delta=0.354$.
Similarly to the case of LSCO, our analysis fits well the
``$\lambda$'' values predicted by others and very different
models~\cite{Shiina90,Gonnelli98,Kordyuk06} (see
table~\ref{tab:table2}). From our results, it is clear that the
phonon coupling mode as a unique source for the behavior of the
critical temperature in BSCCO is only reasonable for the underdoped
case (x=0.12). Nevertheless, the full energy spectrum in the nodal
direction and the consequent emergence of the kink-effect are
reproduced.


\paragraph{Results for YBCO.--} Finally, in Fig.~\ref{Fig:4} we show the results found for Y123
samples. The experimental data were taken from recent work by
Borisenko \emph{et al.}\cite{Borisenko06} To our knowledge, no more
experimental evidence of kinks in the nodal direction is available
for Y123. The value $v_{F_{<}}=1.63eV \cdot${\AA} has been used for
consistency with the experimental results reported by those authors.
The best fit between the experimental data and our model has been
found with the value $\delta=0.365$. It must be noted that the
appearance of a second kink in the underdoped case may not be
allocated in the current model, and requires further theoretical
considerations. Contrary to the properties of the overdoped case,
that is captured by the theory in the full energy spectrum, the
underdoped case is only reproduced at low energies (1st kink close
to the Fermi level). One possibility for upgrading the theory would
be to consider the existence of additional contributions to the
electron self-energy $\Sigma$ which could be related to high orders
of the phonon perturbation, vertex corrections, or inclusive
Coulombian effects. On the other hand, one specific feature of Y123
is that, contrary to the other superconductor cuprates, the
reservoir layers in these materials contain $CuO$ chain layers which
could be contributing significantly to the band energies of the
in-plane electronic structure. This hypotheses has been used with
success by Cucolo \emph{et al.} \cite{Cucolo96} for the
interpretation of tunneling spectra, specific heat and the
ultrasonic attenuation coefficient in both phases of Y123. More
detailed studies on the electronic photoemission spectra of Y-based
copper oxides are required. Unfortunately on the basis of the
existing data, then, it is not possible to favor one or other of
these possibilities. Moreover, in ARPES one should also care about
the residual 3-dimensionality and its effect on photoemission data.
Again, our model fits well the ``$\lambda$'' values predicted by
other works~\cite{Shiina90,Weber88,Heid08} (see
table~\ref{tab:table3}). However, is clear that the phonon coupling
mode as the source for the critical temperature in YBCO samples is
not reasonable. The same conclusion has already been obtained in the
Refs.~\onlinecite{Weber88} and \onlinecite{Heid08}.
%
%
\begin{table}
\caption{\label{tab:table3} Same as tables~\ref{tab:table1}
and~\ref{tab:table2} but for $YBa_{2}Cu_{3}O_{6+x}$ samples. In this
case, $\delta=0.365$}
\begin{ruledtabular}
\begin{tabular}{cccccc}
SC&$x$&Ref.&$\lambda$&$\lambda^{*}$&$T_{c}(\lambda)$  [K]\\
\hline
YBCO   & 0.4  & This\footnotemark[1] & 0.80 & 0.29 & 17.51 \\
       & 0.6  &                      & 0.65 & 0.24 & 9.82 \\
       & 0.85 &                      & 0.50 & 0.18 & 3.39 \\
\hline
       &  --  & [\onlinecite{Shiina90}]\footnotemark[2] & $\sim$3.4 & -- & 91 \\
       &  --  & This  & 3.45 & 1.26 & 84.19 \\
\hline
     & -- & [\onlinecite{Weber88}]\footnotemark[2] & $\sim$0.5 & -- & $\sim$3 \\
     & -- & [\onlinecite{Weber88}]\footnotemark[2] & $\sim$1.3 & -- & $\sim$30 \\
     & -- & This                           & 1.30       & 0.47 & 36.43 \\
\hline
     & -- & [\onlinecite{Heid08}]\footnotemark[3]  & -- & 0.18 - 0.22 & -- \\
     & -- & This                           & 0.49 - 0.60 & 0.18 - 0.22 & $\sim$3.0 - 6.6 \\
\end{tabular}
\end{ruledtabular}
\footnotetext[1]{We allow a margin of error in $\lambda$ of $\sim
\pm 0.3$ as related to the numerical interpolation procedure between theory and experiment.}
\footnotetext[2]{The $\lambda$ values reported
in that reference were obtained so as to fit $T_{c}$ at the indicated values.}
\footnotetext[3]{In Ref.~\onlinecite{Heid08} the parameter
``$\lambda$'' has been obtained from the spectral density
$\alpha^{2}F(\textbf{k},\nu)$ employing the local density
approximation to density functional theory (see text). This value
correspond at the mass-enhancement parameter $\lambda^{*}$}
\end{table}

\section{Concluding remarks}\label{Sec:discussion}


In summary, we have introduced a model that allows to reproduce the
appearance of the ubiquitous nodal kink for a wide set of ARPES
experiments in cuprate superconductors. Our proposal is grounded on
the Migdal-Eliashberg approach for the self-energy of
quasi-particles within the electron-phonon coupling scenario. The
main issue is the proposal of a linear dispersion relation for the
bare band energy, i.e.:
$\varepsilon_{k}=(k-k_{F})v_{F<}(1-\delta\lambda)^{-1}$. $\delta$,
the only free parameter of the theory is a universal property for
each family of cuprates. It has been interpreted as the relation
between the mass-enhancement $\lambda^{*}$ and electron-phonon
coupling $\lambda$ parameters.


An excellent agreement between the theory and the available
collection of experiments is achieved. Our results support the idea
that the phonon coupling mechanism is the main cause of the kink
effect and the so-called universal nodal Fermi velocity, although
its effect in the appearance of the superconducting state and the
high critical temperatures is not clear yet.


For decades, a well-known controversy has arisen on the role of the
parameter ``$\lambda$'' whose values noticeably scatter among
different model calculations. As a central result, our proposal
re-ensembles the ``$\lambda$'' values obtained from different models
and, as a first approximation, solves the controversy through the
relation $\lambda^{*}\cong\delta\lambda$. We emphasize that the
phenomenological parameter $\delta$ (obtained through the analysis
of a wide collection of data) has allowed to go beyond the
conventional Migdal-Eliashberg approach analysis of restricted sets
of experiments.


Our model is directly supported by the ``$\lambda$'' values obtained
in Refs.
\onlinecite{Shiina90,Gonnelli98,Giustino08,Kordyuk06,Heid08,Weber87,Weber88}.
When inserted into the celebrated McMillan's formula these values
indicate that the electron-phonon coupling behind is not necessarily
the only mechanism responsible for the superconducting properties of
all the cuprates. In fact, the critical temperatures that one can
calculate are only at reasonable levels for the case of LSCO
(table~\ref{tab:table1}) and for underdoped BSCCO
(Table~\ref{tab:table2}) where the phonon mechanism dominates.

\section*{Acknowledgement}


The authors want to acknowledge the suggestions of the referees,
that have been of much help for improving the final version of
different parts of this article. This work was partially supported
by Spanish CICyT projects MTM2006-10531 and MAT2005-06279-C03-01.
Contract No. 20101009395-200706 Universidad Nacional de
Colombia-Banco de la República, and Universidad de Zaragoza-Banco
Santander.
%


\end{document}